\documentclass[letterpaper]{IEEEtran}
\usepackage{amsmath, amssymb, amsthm,mathtools, stmaryrd, bm, gensymb,cite,algorithm,algpseudocode,float,subfloat}
\usepackage{subfigure,xcolor}
\usepackage{caption}

\newcommand{\R}[1]{\mathrm{#1}}
\newcommand{\bof}[1]{{\mbox{\boldmath$#1$}}}

\renewcommand{\baselinestretch}{.98}
\begin{document}
	\title{ Rank-one Detector for Kronecker-Structured Constant Modulus Constellations }
	\author{Fazal-E-Asim,~\IEEEmembership{Student~Member,~IEEE}, Andr\'e L. F. de Almeida,~\IEEEmembership{Senior~Member,~IEEE}, Martin Haardt,~\IEEEmembership{Fellow,~IEEE},  Charles C. Cavalcante,~\IEEEmembership{Senior~Member,~IEEE}  and  Josef A. Nossek,~\IEEEmembership{Life~Fellow,~IEEE}
		\thanks{Fazal-E-Asim, Andr\'e L. F. de Almeida, Charles C. Cavalcante and  Josef A. Nossek are with the Department
			of Teleinformatics Engineering, Federal University of Cear\'a, Brazil. Josef A. Nossek is also with the Department of Electrical and Computer Engineering, Technical University of Munich, Germany. Martin Haardt is with the Communications Research Laboratory, Ilmenau University of Technology, Ilmenau, Germany.
			e-mail: \{fazalasim, josef.a.nossek, charles, andre\}@gtel.ufc.br, martin.haardt@tu-ilmenau.de}}  %
	\maketitle
	\begin{abstract}
		To achieve a reliable communication with short data blocks, we propose a novel decoding strategy for Kronecker-structured constant modulus signals that provides low bit error ratios (BERs) especially in the low energy per bit to noise power spectral density ratio $(E_b/N_0)$. The encoder exploits the fact that any $M$-PSK constellation can be factorized as Kronecker products of lower or equal order PSK constellation sets. A construction of two types of schemes is first derived. For such Kronecker-structured schemes, a conceptually simple decoding algorithm is proposed, referred to as Kronecker-RoD (rank-one detector). The decoder is based on a rank-one approximation  of the ``tensorized'' received data block, has a built-in noise rejection capability and a smaller implementation complexity than state-of-the-art detectors. Compared with convolutional codes with hard and soft Viterbi decoding, Kronecker-RoD outperforms the latter in BER performance at same spectral efficiency.
	\end{abstract}
	\begin{IEEEkeywords}
		Kronecker coding, rank-one approximation, LS- Kronecker factorization, normal approximation
		% \LaTeX, paper, template.
	\end{IEEEkeywords}
	\section{Introduction}
	The next generation of cellular communication system promises high data rates, coverage, and responsiveness of communication networks \cite{Ganesh_2014}. Channel coding becomes an important player to reduce the errors introduced by the channel, where low-density parity-check (LDPC), Turbo, and convolutional codes are major candidates \cite{AGoe_2012}.
	In \cite{JPorath_2003}, the authors proposed a multidimensional signal constellation design based on Euclidean space, which is used for spherical and non-spherical codes and has good performance in additive white Gaussian noise (AWGN) channels. The constellation rotation angle coupled with interleaving is optimized in \cite{MNKhormuji_2006} over fading channels for $M$-PSK (phase shift keying) and $M$-QAM (quadrature amplitude modulation). Furthermore, \cite{EBiglieri_1998} explains the impact of coding and modulation for fading channels. The paper also describes the channel characteristics that impact different equalization techniques. In summary, coding schemes that are optimized for the Gaussian channel are likely to be sub-optimal for Rayleigh fading channels.
	\\
	To design efficient symbol detection schemes, a new parallel low complexity iterative detector is proposed in \cite{GHegde_2016} for $M$-PSK constellations using the Lagrangian duality principle while the authors in \cite{Freitas_2019} proposed a least squares (LS)-Kronecker factorization receiver that exploits a cross-coding using tensor space-time coding (TSTC) based on Kronecker products of symbol matrices.
	The receiver algorithm is build on the Kronecker product approximation problem derived in \cite{VanLoan_1992}.
	\\
	In this paper, we introduce a novel encoding model for constant modulus constellations that exploits the fact that any $M$-PSK  constellation  can  be  factorized  as  Kronecker products  of  lower  or  equal  order  PSK  constellation  sets. Based on this property, a construction  of two types of schemes is first derived. Then, a  conceptually  simple  decoding algorithm is  proposed, referred to as Kronecker-RoD (rank-one detector). The detector performs a rank-one approximation of the ''tensorized'' received data block using a tensor power method. The main contributions are summarized as follows.
		\begin{itemize}
			\item  We introduce a novel coded constellation design that represents the encoded $M$-PSK symbols as a Kronecker product of smaller symbol vectors belonging to PSK constellations of smaller or equal cardinality;
			\item To exploit the benefit of such Kronecker coding at the receiver, an efficient detector is proposed that is based on a rank-one approximation of the tensorized received data. The proposed Kronecker-RoD is conceptually simple, operates with short data blocks, and has a smaller implementation complexity than state-of-the-art detectors;
			\item As shown in our numerical results, Kronecker-RoD has a superior BER performance than hard and soft Viterbi decoding at the same spectral efficiency. We further compare our proposed scheme with the theoretical bound given by the normal approximation.
	\end{itemize}
	\textbf{Notation:}
	Scalars are denoted by lower-case italic letters $(a,b,\dots)$, vectors by bold lower-case italic letters $(\bof{a},\bof{b},\dots)$, matrices by bold upper-case italic letters $(\bof{A},\bof{B},\dots)$, tensors are defined by calligraphic upper-case letters $(\mathcal{A},\mathcal{B},\dots)$, $\bof{A}^\R{T}$,$\bof{A}^\R{\ast}$,$\bof{A}^\R{H}$ stand for transpose, conjugate and Hermitian of $\bof{A}$, respectively. The operators $\otimes$ and $\circ$ define the Kronecker and the outer product, respectively. For an $N$th order tensor $\mathcal{Y} \in \mathbb{C}^{L_1\times\dots\times L_N}$, the $n$-mode unfolding of $\mathcal{Y}$ is the matrix $\left[
	\mathcal{Y}\right]_{(n)} = \mathbb{C}^{L_n \times L_1\dots L_{n-1}L_{n+1}\dots L_N}$. $\mathbb{E}[\cdot]$ is the expectation operator and $\bof{I}$ is the identity matrix.
	\begin{figure}[!t]
		\centering
		\subfigure[$\Phi_0\in$ BPSK]
		{\includegraphics[width=1.2in]{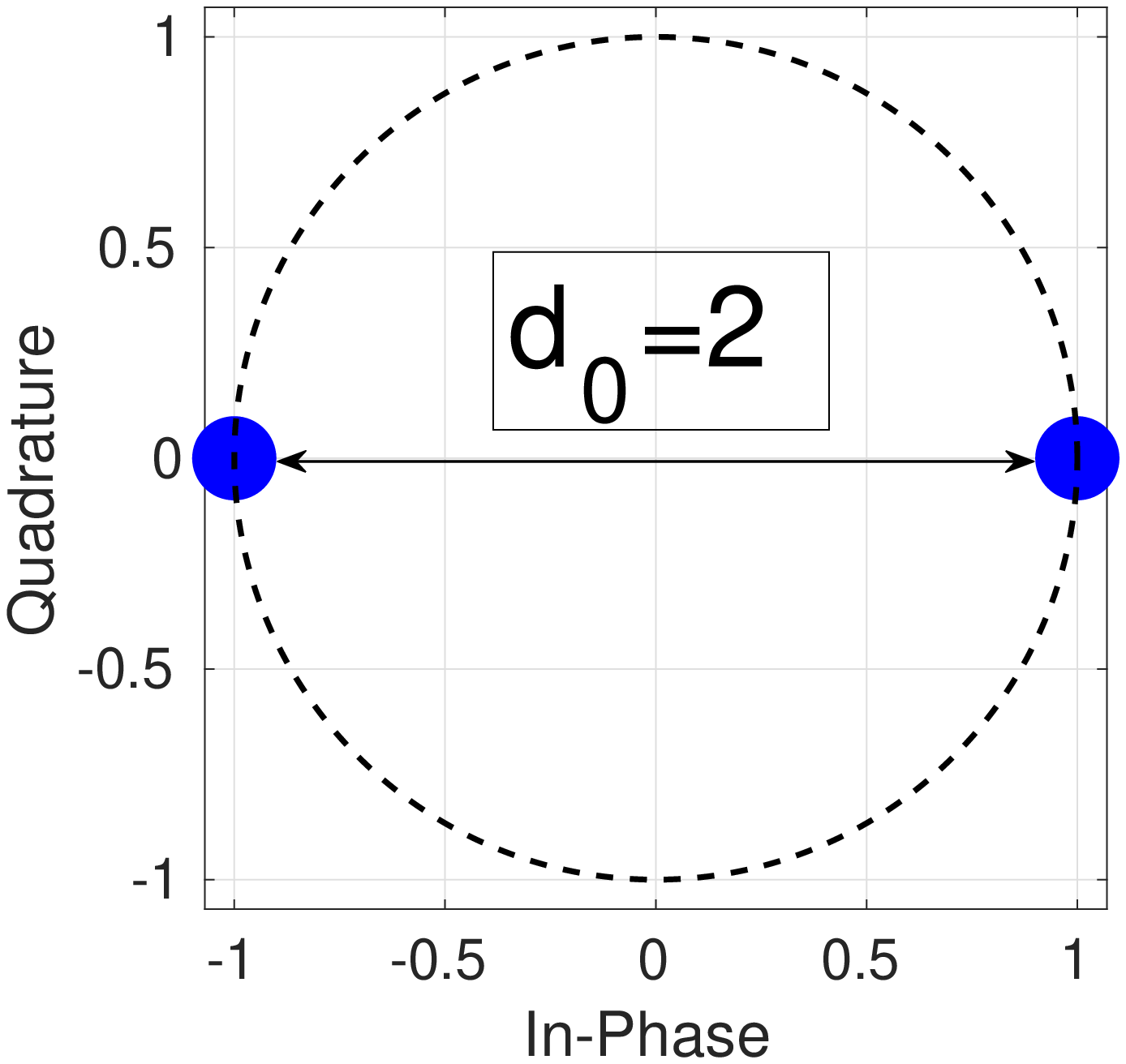}
			\label{fig:second_sub2}
		}
		\subfigure[$\Phi_1$]
		{
			\includegraphics[width=1.2in]{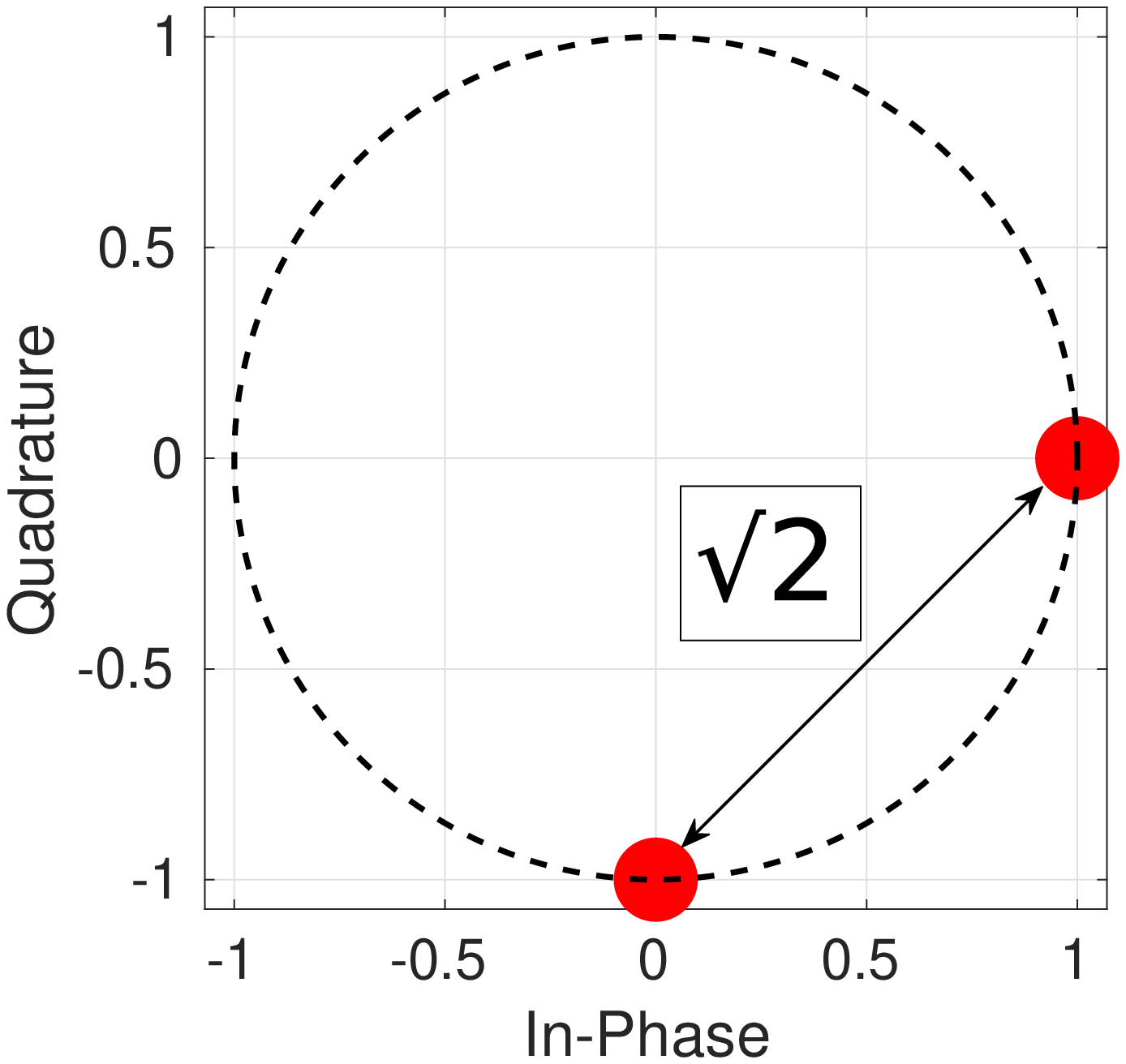}
			\label{fig:third_sub3}
		}
		\\
		\subfigure[$\Phi_2$]
		{
			\includegraphics[width=1.2in]{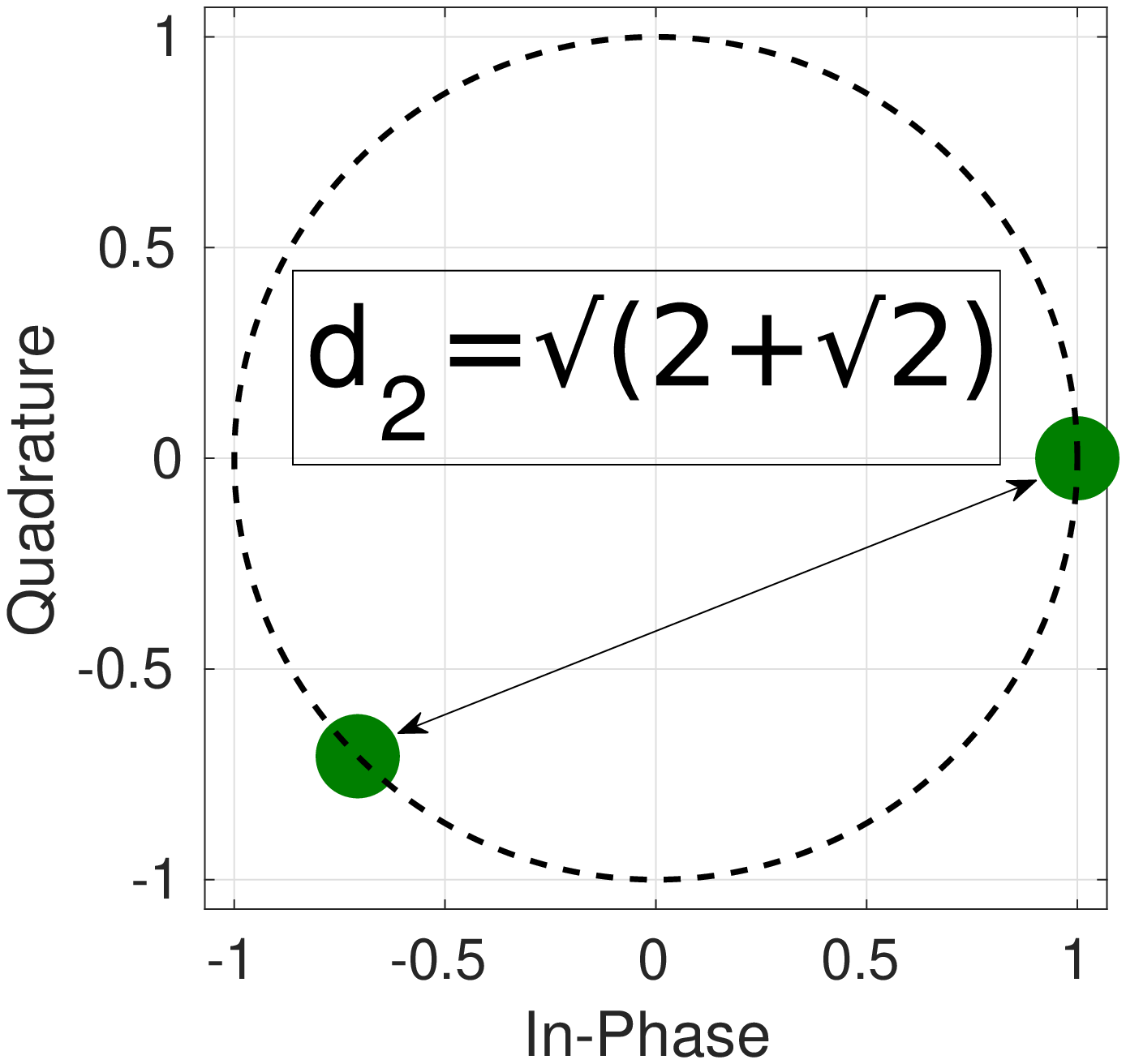}
			\label{fig:fourth_sub4}
		}
		\subfigure[$\Phi = \Phi_0\otimes\Phi_1\otimes \Phi_2$]
		{
			\includegraphics[width=1.2in]{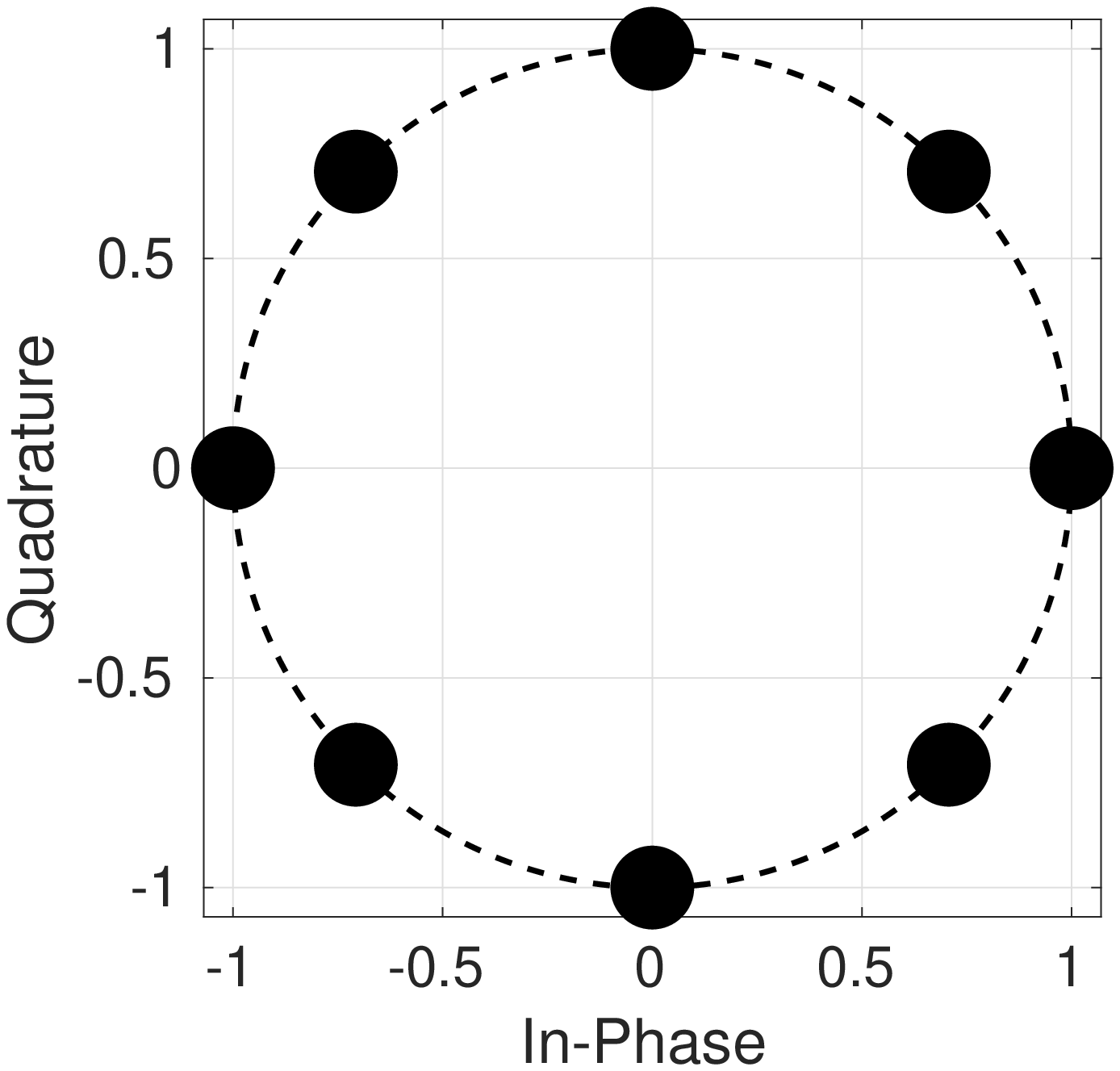}
			\label{fig:first_sub1}
		}
		\caption{Kronecker-structured 8-PSK constellation.}
		\label{fig:8PSKexample}	
	\end{figure}
	\section{Kronecker-Structured Constellation}
	The basic principle behind the Kronecker-structured constellation is stated as follows. Any known $M$-PSK constellation can be factorized into at least $P=\log_2 M$ constellation sets, i.e.,
		\begin{equation}
		\Phi = \Phi_0 \otimes \Phi_1 ,\dots, \otimes \,\Phi_{P-1}  \label{Kron_struct_const}
		\end{equation}
		where $\Phi_0$ is the basis ($M$-PSK) constellation, while $\Phi_{p}, p=1,\dots,P-1$ are the remaining constellation sets, which are either special binary PSK constellations (whose constructions are detailed in the following section) or a combination of conventional PSK constellations of cardinalities $M_1, \ldots, M_{P-1}$, $M_p \leq M$,  $p=1,\dots,P-1$.
	\subsection{Scheme 1} \label{scheme_1}
	In this scheme, the basis constellation set $\Phi_0 \in \left\{1, \quad e^{j\pi}\right\}$ is a BPSK constellation while the remaining $P-1$ constellation sets are binary-PSKs related by a fixed successive rotation of $\frac{\pi}{2^p}$, $p=1,\dots,P-1$, applied to the basis BPSK constellation in the clock wise direction as
		\begin{equation}\label{eq:Phip_scheme1}
		\Phi_{p} \in \left\{\begin{matrix}
		1, \quad e^{j\left(\pi + \frac{\pi}{2^p}\right)}
		\end{matrix}\right\} ,\quad p=1,\dots,P-1.
		\end{equation}
		For illustration purposes, let us consider the case of a $8$-PSK constellation shown in \figurename {~\ref{fig:8PSKexample}}. There is a minimum of $P=\log_2 (8)=3$ binary-PSK constellation sets required to generate the $8$-PSK one. The first constellation $\Phi_0$ is the well-known BPSK one, as shown in \figurename{~\ref{fig:second_sub2}}. The second and third ones, are special binary-PSK constellations $\Phi_1$ and $\Phi_2$ as shown in \figurename{~\ref{fig:third_sub3}} and \figurename{~\ref{fig:fourth_sub4}}, respectively. Note that the Kronecker products involving the three constellations in Figs. \ref{fig:second_sub2}-\ref{fig:fourth_sub4} result in a conventional $8$-PSK constellation, as shown in \figurename{~\ref{fig:first_sub1}. Table \ref{tab:1} provides some examples of Kronecker-structured constellations designed for QPSK and 8-PSK.}
	\subsection{Scheme 2} \label{scheme_2}
	\begin{figure*}[!t]
		\centering
		\subfigure[Kronecker encoding and decoding for SISO system.]
		{
			\includegraphics[width=4.1in]{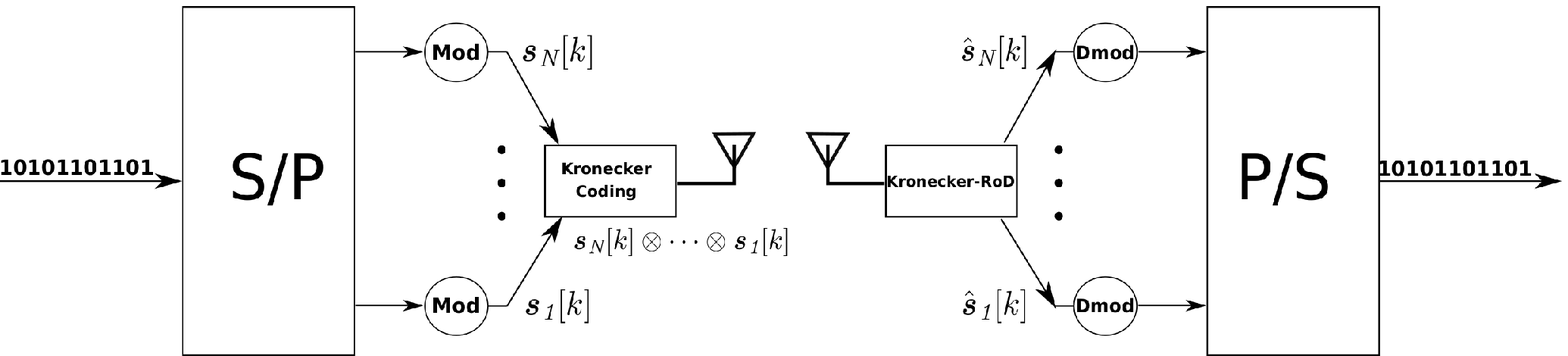}
			\label{fig:systemmodel}
		}
		\subfigure[Implementation of Kronecker-RoD as parallel TMPD branches.]
		{\includegraphics[width=2.8in]{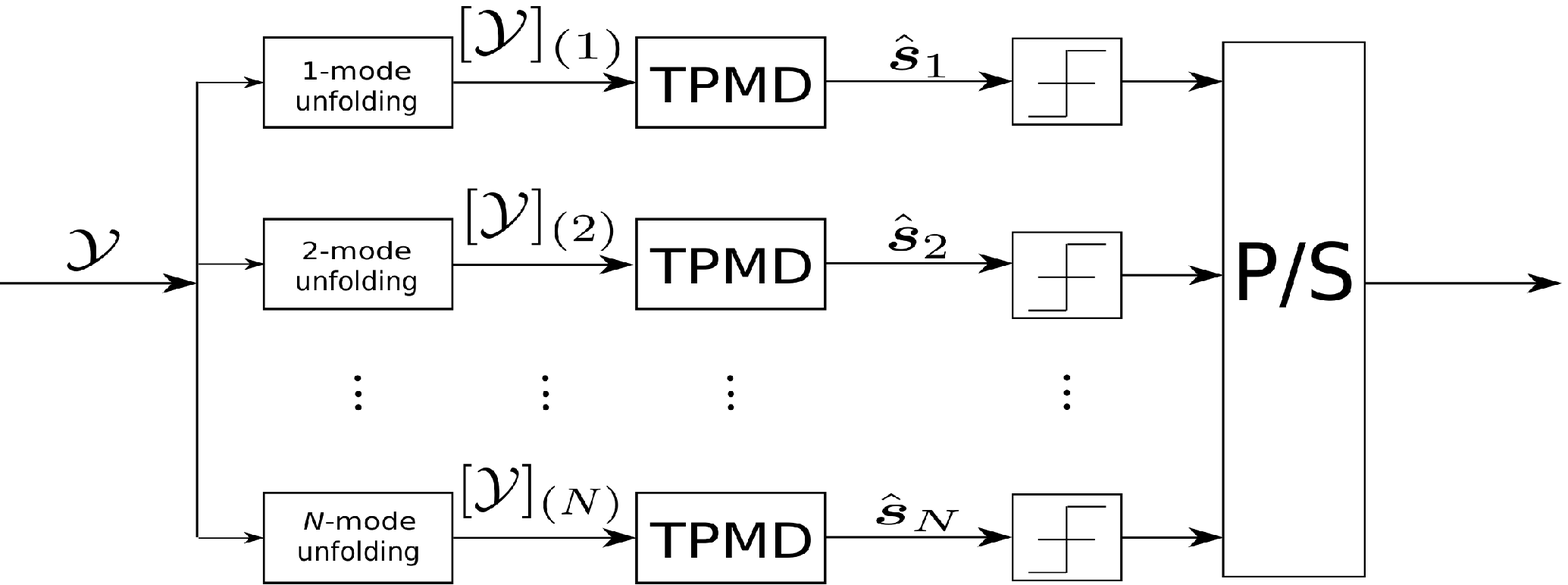}
			\label{fig:powermethod}
		}
		
		\caption{Kronecker encoding and parallel TPMD decoder.}
		\label{fig:SYSTEM MODEL}
	\end{figure*}
	In scheme 2, the basis constellation $\Phi_0$ is taken from any known $M_p\textrm{-PSK}, p=0,\dots,P-1$ constellation and the remaining constellation sets $\Phi_{p}, p=1,\dots,P-1$ are lower or equal orders $M_p$-PSK, $p=1,\dots,P-1$. As an example, a $8$-PSK can be formed as the basis constellation $\Phi_0 \in \textrm{BPSK}$, $\Phi_1 \in \textrm{QPSK}$ and $\Phi_2 \in 8\textrm{-PSK}$. Table \ref{tab:1} provides examples of 8-PSK  Kronecker-structured constellations using different combinations of constellation sets. As a specific case, we can also assume repeated values of $M_p\leq M$, $p=0\ldots P-1$, which means that two or more constellation sets follow the same modulation cardinality, which is at most $M$. 	
		\begin{table}[h]
		\vspace{-2ex}
		\caption{Generation of Kronecker-structured constellations. }
		\label{tab:1}
		\centering
		\begin{tabular}{c|c|c|c|c}
			\bfseries Scheme & \bfseries  $\Phi$ & \bfseries $\Phi_0$ & \bfseries $\Phi_1$ & \bfseries $\Phi_2$  \\
			\hline
			1 & QPSK  & $\left\{1, \, e^{j\pi}\right\}$ & $\left\{1, \, e^{j\left(\pi + \frac{\pi}{2}\right)}\right\}$ & $-$ \\
			1 & 8-PSK  & $\left\{1, \, e^{j\pi}\right\}$ & $\left\{1, \, e^{j\left(\pi + \frac{\pi}{2}\right)}\right\}$ & $\left\{1, \, e^{j\left(\pi + \frac{\pi}{4}\right)}\right\}$ \\
			\hline
			2 & 8-PSK & BPSK & QPSK & 8-PSK  \\
			2 & 8-PSK & BPSK & BPSK & 8-PSK  \\
			2 & 8-PSK & QPSK & QPSK & 8-PSK \\
			2 & 8-PSK & 8-PSK & 8-PSK & 8-PSK
		\end{tabular}
		\vspace{-2ex}
	\end{table}
	\subsection{Probability of error for Kronecker-structured constellation}
	The bit error probability $\text{P}(\Phi_{p})$ for each $p$-th binary-PSK using scheme 1 is given as
	\begin{equation}
	\text{P}(\Phi_p) = Q \left(\frac{d_{\text{min}}^p}{\sqrt{2N_0}}\right),  p=1,\dots,P-1,
	\end{equation}
	where $d_\text{min}^p$ is the minimum Euclidean distance between two binary-PSK symbols of the $\Phi_{p}$ \eqref{eq:Phip_scheme1}. The range of minimum Euclidean distance for scheme 1 is $d_\text{min} \in \left\{\sqrt{2},2\right\}$ as shown in  \figurename{~\ref{fig:8PSKexample}}. For scheme 1, the overall probability of error is dominated by that of the binary constellation set $\Phi_1$ which is the one having the minimum Euclidean distance among the $P-1$ binary constellation sets. This can be seen by inspecting Eq. (\ref{eq:Phip_scheme1}) and also in the example in \figurename{~\ref{fig:8PSKexample}}.
	While, in scheme 2, all the constellations $\Phi_{p}, p=0,\dots,P-1$ involved in the Kronecker product \eqref{Kron_struct_const} are known $M_p$-PSK, consequently the error probability is well known in the literature for every individual constellation set $\Phi_{p} \in M_p\text{-PSK}$ \cite{Proakis2007}. Hence, the overall probability of error is dominated by the $M_p$-PSK, $p=0,\dots,P-1$ having maximum cardinality. 		
	\section{Kronecker-RoD Detector} \label{Kro_RoD}
	\subsection{System Model} \label{sec:model}
	We consider a single-input-single-output (SISO) communication system.  The received vector $\bof{y}[k] \in \mathbb{C}^{L\times1}$ is given by
	\begin{equation}
	\bof{y}[k] = h[k]\bof{x}[k] + \bof{n}[k],    \label{systemmodel}
	\end{equation}
	where $\bof{x}[k] \in \mathbb{C}^{L\times1}$ is the transmitted data drawn from a constant modulus constellation, $h[k]$ represents the fading channel coefficient, $L$ is the length of the transmitted data block, and $\bof{n}[k]\sim \mathcal{C}\mathcal{N}(\bof{0}_{L \times 1},\sigma_n^2\bof{I}_L)$ is the additive white Gaussian noise, where $\sigma_n^2$ denotes its variance. The system model is shown in \figurename{~\ref{fig:systemmodel}}. Let us first give a high-level overview of the main blocks. At the transmitter, the serial bit stream is converted to $N$ parallel streams. At the $n$-th branch, the corresponding binary stream is modulated according to the available constellation sets $\Phi_0,\dots,\Phi_{P-1}$, the design of which are detailed in the previous section for schemes 1 and 2.
	After the modulation, the Kronecker coding is applied and the resulting coded data block is transmitted. At the receiver, the Kronecker-RoD is applied to the received block, which is processed by multiple TPMD branches, as explained in $\R{Section}$ \ref{TPMD}, to estimate the symbol vectors $\hat{\bof{s}}_n[k],n=1,\dots,N$, before demodulation and conversion back to a serial bit stream.	We consider a Kronecker-structured constellation design for the transmitted data, which consists of ``cross-coding'' $N$ symbol vectors as follows
	\begin{equation}
	\bof{x}[k] = \bof{s}_N [k] \otimes\dots \otimes \bof{s}_1 [k] \in \mathbb{C}^{L \times 1},\label{data}
	\end{equation}
	where $\bof{s}_n[k] \in \mathbb{C} ^{L_n \times 1}$ , $n=\{1,\dots,N\}$, and $L = \Pi_{n=1}^N L_n.$
	It is worth mentioning that the same constellation set $\Phi_p$, $\forall p=1,\ldots, P-1$, can be reused multiple times to map $\{\bof{s}_n[k]\}, n=1,\dots,N$. Table \ref{tab:2}, provides some examples of constellation mappings for schemes 1 and 2, assuming $N\geq P$.
	The data rate in bits/symbol period (bit rate) is given as
	\begin{equation}
	\Upsilon = \frac{\sum_{n=1}^{N}L_n \log_2 M_n}{\Pi_{n=0}^{N}L_n}. \label{Gdatarate}
	\end{equation}
	Taking a closer look, we can conclude $\Upsilon_{\text{scheme 1}} \le \Upsilon_{\text{scheme 2}} \le \log_2 M \Upsilon_{\text{scheme 1}}$,
	which shows that scheme 2 always has a higher data rate than scheme 1 by construction. In the special case with $M_0=M_1,\dots=M_{P-1}=M$-PSK constellation, we have  $\Upsilon_{\text{scheme 2}} = \log_2(M) \Upsilon_{\text{scheme 1}}$. Note that the data rates of both schemes are equal when BPSK is used to modulate the $N$ symbol vectors. In addition, assuming the same length for all the symbol vectors, say, $L_n=l$, the data rate boils down to $\frac{N \sum_{n=1}^{N} \log_2M_n}{l^{N}}$. Finally, the code rate $\textit{R}$ for the Kronecker-coding is given by $\frac{\sum_{n=1}^{N}L_n}{\Pi_{n=1}^NL_n}$.
	\begin{table}[t]
		\vspace{-2ex}
		\caption{Relation between $N$ and $P$ for 8-PSK.}
		\label{tab:2}
		\centering
		\begin{tabular}{c|c|c}
			\bfseries Scheme  & \bfseries $(N,P)$ & \bfseries  $\bof{x}[k]$   \\
			\hline
			1   & $(3,3)$ & $\left\{\bof{s}_1,\bof{s}_2,\bof{s}_3\right\} \leftrightarrows \left\{\Phi_0,\Phi_1,\Phi_2\right\}$ \\
			1   & $(5,3)$ & $\left\{\bof{s}_1,\bof{s}_2,\bof{s}_3,\bof{s}_4,\bof{s}_5\right\} \leftrightarrows \left\{\Phi_0,\Phi_0,\Phi_0,\Phi_1,\Phi_2\right\}$ \\
			\hline
			2  & $(2,2)$ & $\left\{\bof{s}_1,\bof{s}_2\right\} \leftrightarrows \left\{\Phi_0,\Phi_1\right\} \in (2,8)$-PSK  \\
			2  & $(3,2)$ & $\left\{\bof{s}_1,\bof{s}_2,\bof{s}_3\right\} \leftrightarrows \left\{\Phi_0,\Phi_1,\Phi_2\right\} \in (2,4,8)$-PSK  \\
			2 & $(3,2)$ & $\left\{\bof{s}_1,\bof{s}_2,\bof{s}_3\right\} \leftrightarrows \left\{\Phi_0,\Phi_1,\Phi_2\right\} \in 8$-PSK \\
		\end{tabular}
		\vspace{-2ex}
	\end{table}
	\subsection{Tensor Power Method Detector} \label{TPMD}
	The received signal after matched filtering is given by
	\begin{equation}
	\hat{\bof{y}}[k] =h^\ast[k] \bof{y}[k] \in \mathbb{C}^{L \times 1}.
	\end{equation}
	For notational convenience, we skip the index $[k]$ in the rest of this paper, since it is irrelevant for the proposed detector.
	From \eqref{systemmodel} and \eqref{data}, the estimation of the transmitted data $\bof{x}$ can be translated into minimizing $ \left \| \hat{\bof{y}} - \bof{s}_N\otimes\dots\otimes\bof{s}_1 \right\|^2_2$ in the least squares sense. The key aspect is the fact that \eqref{data} can be recast as a rank-one $N$-th order tensor $\mathcal{X} = \bof{s}_1\circ \dots \circ\bof{s}_N \in \mathbb{C}^{L_1 \times L_2 \times \dots \times L_N}$ so that the optimization problem can be reformulated as a tensor rank-one approximation problem
	\begin{equation}
	\min _{\bof{s}_1,\dots,\bof{s}_N} \left\| \hat{\mathcal{Y}} - \bof{s}_1\circ\dots\circ\bof{s}_N \right\|^2_\R{F}, \label{tensorcost}
	\end{equation}
	where $\hat{\mathcal{Y}} \in \mathbb{C}^{L_1 \times L_2 \times \dots \times L_N} $ is the $N$-th order ``tensorized'' received data tensor. Indeed, since $\mathcal{Y}$ is a scaled version of $\mathcal{X}$ corrupted by additive Gaussian noise, estimating the transmitted symbols $\bof{s}_1,\dots,\bof{s}_N$ translates into finding a rank-one approximation to $\mathcal{Y}$. It is well known that a solution to this problem is provided by the truncated higher-order singular value decomposition (HOSVD)\cite{DeLathauwer_2000,Kolda_2009}. An efficient way to compute this truncated rank-one HOSVD is the tensor power method \cite{Lathauwer1995HigherorderPM_1995,PRegalia_2000,Golub}. As shown in \cite{Golub}, minimizing \eqref{tensorcost} is equivalent to maximizing the tensor Rayleigh quotient
	\begin{equation}
	T({\bof{s}_1,\dots,\bof{s}_N})   = \frac{\left| \left(\bof{s}_N \otimes \dots \otimes\bof{s}_1\right)^\R{T} \R{vec} \left(\hat{\mathcal{Y}}\right) \right|}{\|\bof{s}_1\|_2 \dots \|\bof{s}_N\|_2}.
	\end{equation}
	 For simplicity, let us define the following Gramian
		\begin{equation}
		\bof{A}_n = [\hat{\mathcal{Y}}]_{(n)} [\hat{\mathcal{Y}}]_{(n)}^\R{H} \in \mathbb{C}^{L_n \times L_n} \label{Gramian}
		\end{equation}	
		where $n = 1,\dots,N$. Note that the $n$-mode unfolding of the received tensor $\hat{\mathcal{Y}}$ can be expressed as
		\begin{equation}
		[\hat{\mathcal{Y}}]_{(n)} = \left |h \right|^2 \bof{s}_n \left(\otimes_{i\neq n} \bof{s}_i^\R{T}\right) + [\mathcal{N}]_{(n)}. \label{n_th_mode_unfolding}
		\end{equation}
		Assuming that the transmitted symbol vectors and noise are mutually uncorrelated, combining \eqref{Gramian} and \eqref{n_th_mode_unfolding} yields
		\begin{align}
		\bof{A}_n &= \left |h \right|^4 \bof{s}_n \underbrace{\left(\otimes_{i\neq n}\bof{s}_i^\R{T}\right)\left(\otimes_{i\neq n}\bof{s}_i^\ast\right)}_\gamma \bof{s}_n^\R{H} + \sigma_n^2\bof{I}, \\
		&= \left |h \right|^4 \gamma \bof{s}_n \bof{s}_n^\R{H} + \sigma_n^2\bof{I},
		\end{align}
		which shows that the $n$-th Gramian is a rank-one matrix corrupted by additive noise. We propose a tensor power method detector (TPMD) based on \cite{Lathauwer1995HigherorderPM_1995,PRegalia_2000,Golub}, where an estimate $\hat{\bof{s}}_n, n = \{1,\dots,N\}$ is found from the dominant left singular vectors $\bof{u}_n \in \mathbb{C}^{L_n \times 1}$ (up to scaling) of the $n$-mode unfolding $\left[\mathcal{Y}\right]_{(n)}$ of the received data tensor $\mathcal{Y}, n = 1,\dots,N$.
	The TPMD starts from the initializations $\bof{u}_1^{(0)},\dots,\bof{u}_N^{(0)}$ of the $N$ symbol vectors, which are randomly drawn from the known alphabets that constitute the Kronecker-structured constellation, which are defined by the constellation sets $\Phi_0, \ldots, \Phi_{P-1}$. At the $j$-th iteration, a new estimate is found by $\bof{u}_n^{(j)} = \bof{A}_n \bof{u}_n^{(j-1)}$. Note that the  number $N$ of TPMD branches corresponds to the $N$ symbols vectors $\bof{s}_n[k]$ that are Kronecker-coded using \eqref{data} as shown in \figurename{~\ref{fig:powermethod}}.
	At the end of the $j$-th iteration, a normalized squared error is calculated as $e_j = \frac{\| \bof{u}_n^{(j)} - \bof{u}_n^{(j-1)} \|^2}{\|\bof{u}_n^{(j)}\|^2}$. The convergence is declared if  $e_j \le 10^{-6}$.
		The TPMD algorithm is summarized in Algorithm \ref{algorithm1}.
		On average, the TPMD takes less than 10 iterations for convergence, although our extensive simulations have shown that in the majority of the cases, the convergence is achieved within 3-4 iterations only.
	Due to the unknown norm of $\bof{u}_n$, we have $\hat{\bof{s}}_n = \beta_n \bof{s}_n , n= \{n = 1,\dots,N\}$, where $\beta_n$ is a scaling factor to be determined before demodulation. This factor can thus be calculated by using prior knowledge of the first element of $\bof{u}_n$ and $\bof{s}_n$ as $\beta_n = \frac{\bof{u}_n[1]}{\bof{s}_n[1]}$ (one single training symbol is enough for this purpose). The final estimate of the $n$-th symbol vector is then obtained as $\hat{\bof{s}}_n = \frac{\bof{u}_n}{\beta_n}$.	
	Since the estimation of $\hat{\bof{s}}_1,\dots,\hat{\bof{s}}_N$ can be carried out independently, the $N$ branches of the  Kronecker-RoD shown in \figurename{~\ref{fig:powermethod}} can be executed in parallel. As far as the slicing procedure is concerned, once the convergence is achieved, the estimated symbols at the TPMD branches are projected onto their respective known alphabet sets $\Phi_0, \dots, \Phi_{P-1}$.
		The complexity of the TPMD is that of executing a matrix-vector product in line \ref{line4} of Algorithm \ref{algorithm1}, which is given by $\mathcal{O}\left(8JL_n^N\right)$. The overall complexity of the Kronecker-RoD is $\mathcal{O}(8NJL_n^N)$.
	\begin{algorithm}[t]
		\caption{Tensor Power Method Detector}\label{algorithm1}
		\begin{algorithmic}[1]
			\Procedure{TPMD}{$\hat{\mathcal{Y}}$}
			%		\State $\hat{\mathcal{Y}}\gets \hat{\bof{y}}$
			\For{ $n = 1$ to $N$ }
			\State Initialize\texttt{ $\bof{u}_n^{(0)}$ } with random $M$-PSK symbols
			\For{ $j = 1$ to $J$ } %\Comment{$J$  iterations}
			\State \texttt{ $\bof{u}_n^{(j)} = \bof{A}_n\bof{u}_n^{(j-1)}$ }  \label{line4}
			%		\State \texttt{ $\mathbf{u}_1 \gets \mathbf{u}_1/\neuclid{\mathbf{u}_1}$ }  \label{al_line_5}\Comment{Left singular vector}
			\EndFor
			\State \texttt{$\beta_n = \frac{\bof{u}^{(J)}_n[1]}{\bof{s}_n[1]}$ \textrm{and} $\hat{\bof{s}}_n = \frac{\bof{u}^{(J)}_n}{\beta_n}$}
			%		\State \texttt{$\hat{\bof{s}}_n = \frac{\bof{u}_n}{\beta_n}$}
			\State \textbf{return} $\hat{\bof{s}}_n$
			\EndFor \;\; \textbf{$N$-modes of $\hat{\mathcal{Y}}$}
			\EndProcedure		
		\end{algorithmic}
	\end{algorithm}
	\section{Numerical Results}
\begin{figure}[t]
		\centering
		\subfigure[AWGN channel.]
		{\includegraphics[scale=0.32]{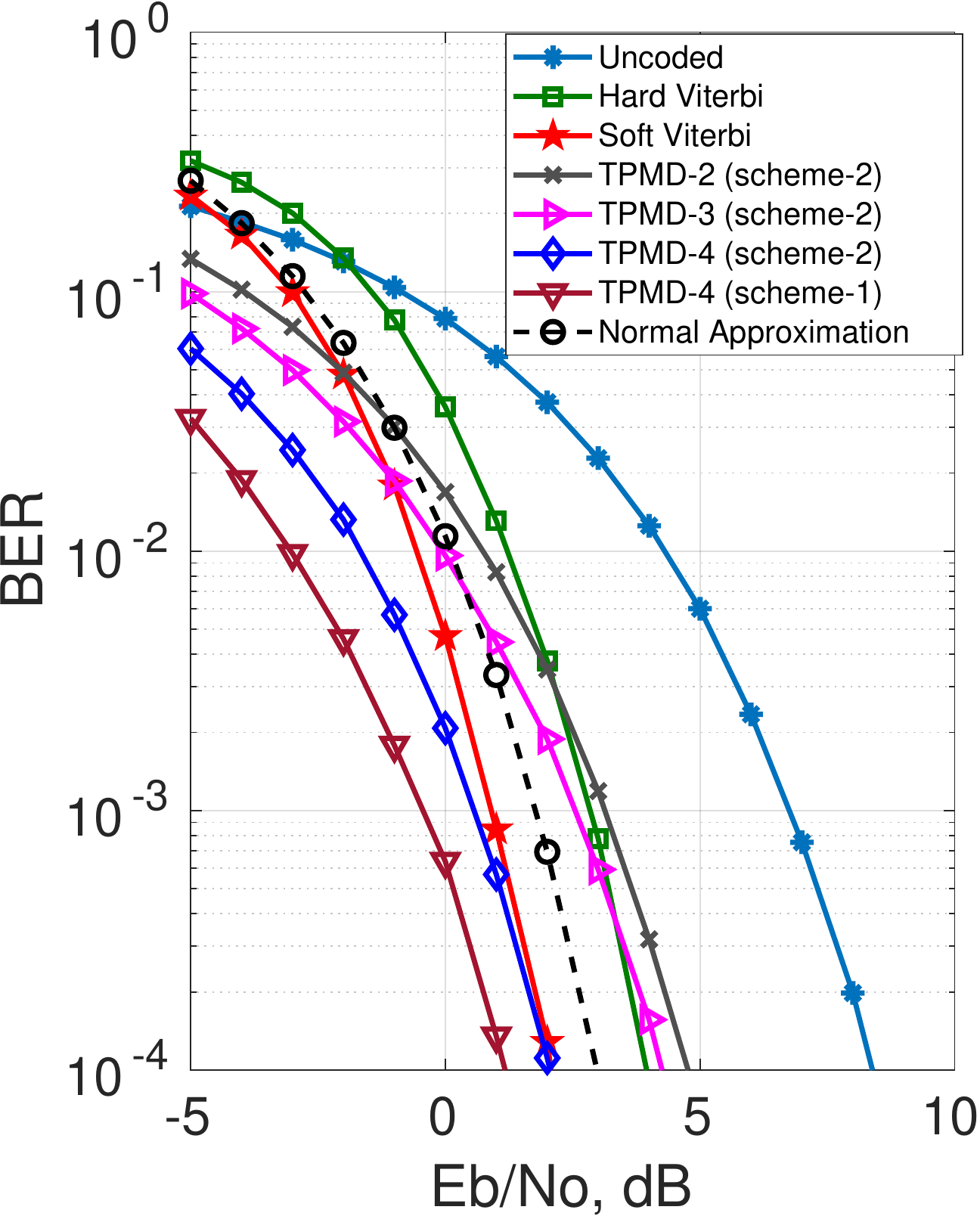}
			\label{fig:qpsk_awgn}
		}
		\subfigure[Flat-fading channel.]
		{\includegraphics[scale=0.32]{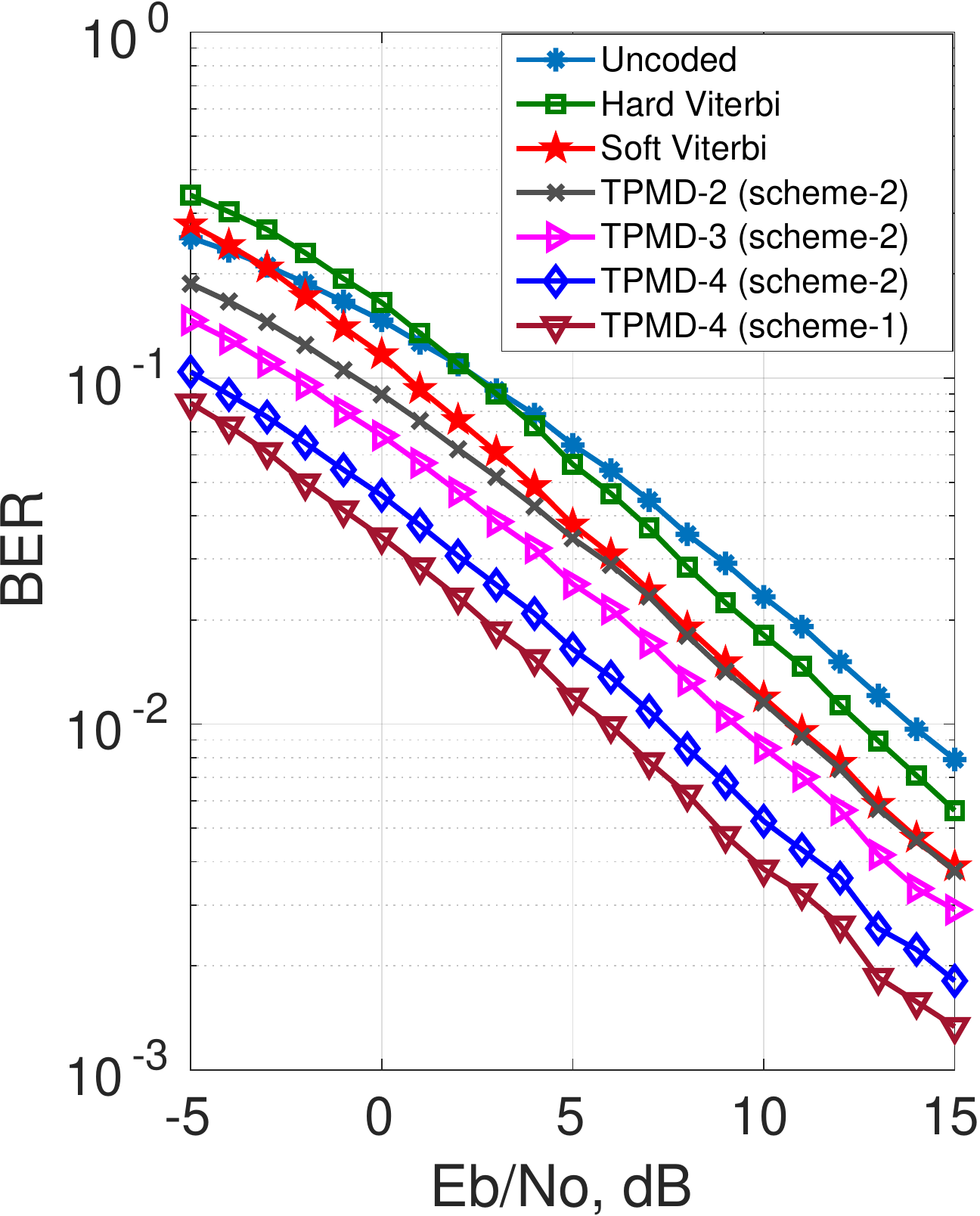}
			\label{fig:qpsk_ray}
		}
		\caption{BER performance comparison of Kronecker-RoD.} \label{fig:BER_performance}
	\end{figure}

	In this section, we evaluate the BER performance of the Kronecker-RoD using TPMD-$N$. To ensure a fair comparison, the best half-rate convolutional encoder is used \cite{AGraell_2004}, with constraint length $K=3$ having generators in the octal form as (5,7). We further compare our proposed method to the normal approximation, which is a theoretical bound designed for short block coding schemes \cite{Polyanskiy_2010,Durisi_2016}.
	Our first experiments consider an AWGN channel as a reference, followed by results under the assumption of a Rayleigh fading channel. In \figurename{~\ref{fig:qpsk_awgn}}, we compare the performance of the different schemes for 4-PSK modulation, resulting in half-rate transmission for all of them. Firstly, note that by increasing the number $N$ of Kronecker-encoded symbol vectors in \eqref{data}, and consequently, the number $N$ of parallel TPMD branches (see \figurename{~\ref{fig:powermethod}}), the BER performance is improved. For a target BER of $10^{-2}$, TPMD-4 (scheme 2) provides an $E_b/N_0$ gain of approximately 2 dB over TPMD-2 (scheme 2). Such an improvement comes from the enhanced noise rejection capability of Kronecker-RoD as $N$ (i.e., the order of the tensorized transmitted/received data block) increases. We can also observe that TPMD-4 (scheme 2) outperforms the soft Viterbi decoder for the lower $E_b/N_0$ range, and significantly outperforms the hard Viterbi decoder. The TPMD-4 (scheme 1) outperforms all approaches, which is an expected result because of the lower error probability of the binary-PSK constellations of scheme 1 compared to those of scheme 2. As far as the comparison with the normal approximation bound is concerned, \figurename{~\ref{fig:qpsk_awgn}} includes results of the Kronecker-RoD using 4-PSK constellations. We can see that all TPMD schemes outperform the normal approximation in the low $E_b/N_0$ regime. In particular, TPMD-4 (operating under both schemes 1 and 2) outperforms the normal approximation in the whole $E_b/N_0$ range. Such a performance gain comes from the enhanced noise rejection capability of TPMD, as the number of branches is increased. In addition, note that the normal approximation assumes i.i.d. Gaussian distributed symbols while the proposed Kronecker-structured constellation introduces a controlled correlation
		on the transmitted symbols that is efficiently exploited by the TPMD receiver to extract the coding gains and reduce the additive noise. \\
Next, a flat Rayleigh fading channel is taken into account as shown in \eqref{systemmodel}. At each run, the channel coefficient is drawn from a zero mean unit variance complex Gaussian distribution. The results are shown in \figurename{~ \ref{fig:qpsk_ray}}, for 4-PSK modulation. Note that the Kronecker-RoD with TPMD-3 (scheme 2) and TPMD-4 (scheme 2) outperforms the hard and soft decision Viterbi decoder for the whole Eb/No range. In particular, TPMD-4 (scheme 2) provides lower error rates than soft Viterbi for lower Eb/No levels. For instance, considering a target BER of $10^{-2}$, Kronecker-RoD with three branches (TPMD-3) provides an $E_b/N_0$ gain of 2 dB over soft Viterbi. Using four branches (TPMD-4) offers a 4 dB gain in terms of $E_b/N_0$ over soft Viterbi, which is a remarkable result. The TPMD-4 (scheme 1) shows a better performance, as expected.
A final remark is related to the decoding delay. For the Kronecker-RoD, the decoding delay is determined by the size of the transmitted symbol block which is $L=16$. Note that, for the Viterbi decoder, according to \cite{Viterbi_1967,BMoision_2008}, the decoding delay is five times the constraint length, i.e., $5K = 5(3)=15$.
	
\section{Conclusion}
In this paper, we have proposed a novel encoding and decoding strategy for constant modulus constellations in SISO communication systems that exploit the Kronecker structure of the cross-coded transmitted symbol vectors. The Kronecker-RoD has a conceptually simple parallel implementation and has a superior performance than competing detectors at the same spectral efficiency. Perspectives include the generalization of the Kronecker-RoD to space-time modulation for multiple-input-multiple-output (MIMO) systems.

\bibliographystyle{IEEEtran}
%	\bibliography{kroncoding}

\end{document}